\documentclass{article}%
\usepackage{amsmath}
\usepackage{graphicx}%
\usepackage{amsfonts}%
\usepackage{amssymb}
\begin{document}

\title{Metric-Field Approach to Gravitation and the Problem of the Universe Acceleration}
\author{Leonid V. Verozub\\Kharkov National University\\verozub@gravit.kharkov.ua}
\date{}
\maketitle


\section{Introduction}

The geometrical properties of space-time can be described only by means of
measuring instruments. At the same time, the description of the properties of
measuring instruments, strictly speaking, requires the knowledge of space-time
geometry. One of the implications of it is that geometrical properties of
space and time have no experimentally verifiable significance by themselves
but only within the aggregate ''geometry + measuring instruments''. We got
aware of it owing to Poincar\'{e} \cite{Poincare}. It is a development of the
Berkley - Leibnitz - Mach idea about relativity of space-time properties,
which is an alternative to the well known Newtonian approach. 

If we proceed from the conception of relativity of space-time, we assume that
there is no way of quantitative description of physical phenomena other than
attributing them to a certain frame of reference which, in itself, is a
physical device for space and time measurements. But then the relativity of
the geometrical properties of space and time mentioned above is nothing else
but relativity of space-time geometry with respect to the frame of reference
being used.\footnote{There is an important difference between a frame of
reference (as a physical device) and a coordinate system (as the way to
parameterize points of space-time) \cite{Rodichev}}. Thus, it should be
assumed that the concept of the reference frame as a physical object, whose
properties are given and are independent of the properties of space and time,
is approximate, and only the aggregate ''frame of reference + space-time
geometry'' has a sense. The Einstein theory of gravitation demonstrates
relativity of space-time with respect to distribution of matter. However,
space-time relativity with respect to measurement instruments hitherto has not
been realized in physical theory. An attempt to show that there is also
space-time relativity to the used reference frames has been undertaken for the
first time in \cite{Verozub81}, \cite{Verozub95}.

\section{Fundamental Metric Form in NIFRs.}

At present we do not know how the space-time geometry in inertial frames of
reference (IFRs) is connected with the frames properties. Under the
circumstances, we simply postulate (according to special relativity) that
space - time in IFRs is pseudo-Euclidean. Next, we find a space-time metric
differential form in noninertial frames of reference (NIFRs) from the
viewpoint of an observer in the NIFR who proceeds from the relativity of space
and time in the Berkley - Leibnitz - Mach - Poincar\'{e} (BLMP) sense.

By a noninertial frame of reference we mean the frame, whose body of reference
is formed by the point masses moving in the IFR under the effect of a given
force field. It would be a mistake to identify ''a priori'' a transition from
the IFR to the NIFR with the transformation of coordinates related to the
frames. If we act in such a way, we already assume that the properties of
space-time in both frames are identical. However, for an observer in the NIFR,
who proceeds from the relativity of space and time in the BLMP sense,
space-time geometry is not given ''a priori'' and must be ascertained from the
analysis of experimental data. We shall suppose that the reference body (RB)
of the IFR or NIFR is formed by the identical point masses $m_{p}$. If the
observer is at rest in one of the frames, his world line will coincide with
the world line of some point of the reference body. It is obvious to the
observer in the IFR that the accelerations of the point masses forming the
reference body are equal to zero. Of course, this fact occurs in relativistic
sense too. Let $d\eta$ and $ds$ be denote the differential metric forms in the
IFR and NIFR. Then, if $\nu^{\alpha}=dx^{\alpha}/d\eta$ is the 4- velocity
vector of the point masses forming the reference body of the IFR, the absolute
derivative of the vector $\nu^{\alpha}$ is equal to zero,i.e.
\begin{equation}
D\nu^{\alpha}/d\eta=0.\label{AbsAcelIFR}%
\end{equation}

From the viewpoint of an observer in a NIFR who proceeds from the relativity
of space-time in the BLM sense the points of his body reference are the points
of his physical space, they are not exposed any forces. Consequently, their
accelerations are equal to zero both in nonrelativistic and relativistic
sense. This means that 4-velocity $\zeta^{\alpha}=dx^{\alpha}/ds$ of the point
masses forming the reference body of the NIFR obeys equality

\bigskip%
\begin{equation}
D\zeta^{\alpha}/ds=0 \label{AbsAcelNIFR}%
\end{equation}
In other words, since for the observer in the IFR according to eq.
(\ref{AbsAcelIFR}) world lines of the IFR points are geodesic lines, then for
the observer in the NIFR world lines of the NIFR points of the reference body
are also geodesic lines in his space-time, which can be expressed by eq.
(\ref{AbsAcelNIFR}).

This fact leads to important consequences. The differential equations of these
world lines at the same time are the Lagrange equations of motion of the NIFR
RB points. The Lagrange equations, describing the motion of the identical RB
point masses in the IFR, can be obtained from the Lagrange action $S$ by the
principle of least action. Therefore, the equations of the geodesic lines can
be obtained from the differential metric form $ds=k\ dS(x,dx)$, where $k$ is
the constant, $dSdt=L(x,\overset{.}{x})dt$ and $L$ is the Lagrange function.
The constant $k=-(m_{p}c)^{-1}$, as it follows from the analysis of the case
when the frame of reference is inertial. Thus, if we proceed from relativity
of space and time in the BLMP sense, the differential metric form of
space-time in the NIFR can be expected to have the following form
\cite{Verozub81}, \cite{Verozub95}.
\begin{equation}
ds=-(m_{p}c)^{-1}\;dS(x,dx).\label{Myds}%
\end{equation}
So, the properties of space-time in the NIFR are entirely determined by the
properties of the used frame in accordance with the idea of relativity of
space and time in the BLMP sense. 

Consider two examples of the NIFR.

1.The reference body is formed by noninteracting electric charges moving in a
constant homogeneous electric field $\mathcal{E}$. The motion of the charges
is described in Cartesian coordinates by the Lagrangian
\begin{equation}
L=-m_{p}c^{2}\ (1-V^{2}/c^{2})^{1/2}+\mathcal{E}~e~x,\label{Vr_lagrframe1}%
\end{equation}
where $V$ is the speed of the charges. According to eq. (\ref{Vr_lagrframe1})
the space - time metric differential form in this frame is given by
\begin{equation}
ds=d\eta-(wx/c^{2})dx^{0},\label{Vr_5/6}%
\end{equation}
where $d\eta=(c^{2}dt^{2}-dx^{2}-dy^{2}-dz^{2})^{1/2}$, is the metric
differential form of the pseudo - Euclidean space - time in the IFR and
$w=e\mathcal{E}/m$ is the acceleration of the charges.

2. The reference body consists of noninteracting electric charges in a
constant homogeneous magnetic field $\mathcal{H}$ directed along the axis $z$.
The Lagrangian describing the motion of the particles can be written as
follows
\begin{equation}
L=-m_{p}c^{2}(1-V^{2}/c^{2})^{1/2}-(m_{p}\Omega_{0}/2)(\overset{\cdot}%
{x}y-x\overset{\cdot}{y}),\label{Vr_5/7}%
\end{equation}
where $\overset{\cdot}{x}=dx/dt$, $\overset{\cdot}{y}=dx/dt$ and $\Omega
_{0}=e\mathcal{H}/2mc$. The points of such a system rotate in the plane $xy$
around the axis $z$ with the angular frequency $\omega=\Omega_{0}%
[1+(\Omega_{0}r/c)^{2}]^{-1/2}$ , where $r=(x^{2}+y^{2})^{1/2}$. The linear
velocities of the BR points tend to $c$ when $r\rightarrow\infty$. For the
given NIFR
\begin{equation}
ds=d\eta+(\Omega_{0}/2c)\ (ydx-xdy).\label{6/8}%
\end{equation}
In the above NIFR $ds$ is of the form $ds=\mathcal{F}(x,dx)$ where
$\mathcal{F}(x,dx)=d\eta+f_{\alpha}(x)dx^{\alpha},$ $f_{\alpha}$ is a vector
field. The function $\mathcal{F}$ is a homogeneous of the first degree in
$dx^{\alpha}$. Therefore, generally speaking, the space-time in NIFR is
Finslerian \cite{Rund} with the sign - indefinite differential metric form.

One of the consequences of the above result is a natural explanation of the
Sagnac effect and the fact of the existance of the inertial forces in NIFRs
\cite{Verozub81}, \cite{Verozub95}.

\section{Experimental Verification}

A clock, which is in a NIFR at rest, is unaffected by acceleration in space -
time of the frame. The change in rate of the ideal clock is a real consequence
of the difference between the space - time metrics in the IFR and NIFR. It is
given by the factor $\sigma=ds/d\eta$ from the equation $ds=\sigma d\eta$. For
the rotating with the angular velocity $\Omega$ disk of the radius $R$ \ the
factor $\sigma=1-\Omega^{2}R^{2}/2c^{2}$ which gives rise to the observed red
shift in the well known Pound - Rebka - Snider experiments.

Another experimentally verifiable consequence of the above theory is some
difference between the inertial mass $m_{p}^{eq}$ of a body on the Earth's
equator and the mass $m_{p}^{pol}$ of the same body on the pole. It is given
by
\begin{equation}
(m_{p}^{eq}-m_{p}^{pol})/m_{p}^{pol}=1.2\cdot10^{-12}%
\end{equation}
The dependence of the inertial mass of particles on the Earth's longitude can
be observed by the M$\overset{..}{\text{o}}$ssbauer effect. Indeed, the change
$\Delta\lambda$ in the wave length $\lambda$ at the Compton scattering on
particles of the masses $m_{p}$ is proportional to $m_{p}^{-1}$. If this value
is measured for gamma - quantums with the help of the M$\overset{..}{\text{o}%
}$ssbauer effect at a fixed scattering angle, then after transporting the
measuring device from the longitude $\varphi_{1}$ to the longitude
$\varphi_{2}$ we have
\begin{equation}
\frac{(\Delta\lambda)_{\varphi_{1}}^{-1}-(\Delta\lambda)_{\varphi_{2}}^{-1}%
}{(\Delta\lambda)_{\varphi_{1}}^{-1}}=\Theta\ [cos^{2}(\varphi_{1}%
)-cos^{2}(\varphi_{2})],
\end{equation}
where $\Theta=1.2\cdot10^{-12}$.

\section{Gravitation in Inertial and Proper Reference Frames}

Consider a frame of reference whose reference body is formed by identical
material points $m_{p}$ moving under the effect of the field $\psi
_{\alpha\beta}$. These frames will be called the proper frames of reference
(PRFs) of the given field. Any observer, located in the PFR at rest, moves in
space-time of this frame along the geodesic line of his space-time. This
implies that the space-time metric differential form in the NIFR is given by
eq. (\ref{Myds}) where $S$ is the action describing in a IFR the motion of
particles forming the reference body of the NIFR. Now suppose, following
Thirring \cite{Thirring}, that in pseudo- Euclidean space-time gravitation can
be described as a tensor field $\psi_{\alpha\beta}(x)$, and the Lagrangian
describing motion of a test particle with the mass $m_{p}$ is of the form
\begin{equation}
L=-m_{p}c[g_{\alpha\beta}(\psi)\overset{\cdot}{x}^{\alpha}x^{\beta}%
]^{1/2},\label{Vr_LagrangianThirr}%
\end{equation}
where $\overset{\cdot}{x}^{\alpha}=dx^{\alpha}/dt$ and $g_{\alpha\beta}$ is
the symmetric tensor whose components are the function of $\psi_{\alpha\beta}%
$. Then, according to (\ref{Myds}) the space-time metric differential form in
the PFR is given by
\begin{equation}
ds^{2}=g_{\alpha\beta}(\psi)\;dx^{\alpha}\;dx^{\beta}%
\end{equation}
Thus, the space-time in the PFR is a Riemannian with the curvature other than
zero. Viewed by an observer in the IFR, the motion of the test particle
forming the reference body of the PFR is affected by the force field
$\psi_{\alpha\beta}$. But the observer located in the PRF will not observe the
force properties of the field $\psi_{\alpha\beta}$ since he moves in
space-time of the PRF along the geodesic line. For him the presence of the
field $\psi_{\alpha\beta}$ will be displayed in another way --- as space-time
curvature differing from zero in these frames, e.g. as a deviation of the
world lines of the neighbouring points of the reference body. For example,
when studying the Earth's gravity, a frame of reference fixed to the Earth can
be considered as an inertial frame if the forces of inertia are ignored. An
observer located in this frame can consider motion of the particles forming
the PRF reference body in flat space-time on the basis of eq.
(\ref{Vr_LagrangianThirr}) without running into contradiction with
experiments. However, the observer in the PFR (in a comoving frame for free
falling particles) does not find the Earth's gravity as some force field. If
he proceeds from the relativity of space-time, he believes that point
particles, forming the reference body of his reference frame, are the point of
his physical space. They are not affected by a force field and, therefore,
their accelerations in his space-time are equal to zero. In spite of that, he
observes a change in the relative distances of these particles. Such an
experimental fact has apparently the only explanation as non-relativistic
display of the deviations of the geodesic lines caused by space-time
curvature. So, we observe an important fact that only in proper frames of
reference we have an evidence for gravitation identification with space-time
curvature. 

Thus, we arrive at the following hypothesis. In inertial frames of reference,
where space-time is pseudo-Euclidean, gravitation is a field $\psi
_{\alpha\beta}$. In the proper frames of reference of the field $\psi
_{\alpha\beta}$, where space-time is Riemannian, gravitation manifests itself
as curvature of space-time and must be described completely by the geometrical
properties of the latter.

Of course, eq. (\ref{Myds}) refers to any classical field. For instance,
space-time in the PRF of an electromagnetic field is Finslerian. However,
since $ds$ depends on the mass $m_{p}$ and charge $e$ of the point masses
forming the reference body, this fact is not of great significance.

\section{Geodesic-invariant equations of gravitation}

Gravitational equations should be some kind of differential equations for the
function $\psi_{\alpha\beta}$ or $g_{\alpha\beta}\;(\psi)$, which are
invariant under a certain kind of gauge transformations of the potentials
$\psi_{\alpha\beta}$. Since $g_{\alpha\beta}=g_{\alpha\beta}\;(\psi)$, the
Einstein equations are the equations both for $g_{\alpha\beta}$ and for
$\psi_{\alpha\beta}$. Under the transformation $\psi_{\alpha\beta}%
\rightarrow\overline{\psi}_{\alpha\beta}$ the quantities $g_{\alpha\beta
}\;(\psi)$ undergo some transformations too and, as a consequence, the
equations of the test particle motion resulting from eq.
(\ref{Vr_LagrangianThirr}) and the Einstein's equations do not remain
invariant. The equations of motion resulting from eq.
(\ref{Vr_LagrangianThirr}) are at the same time the equations of a geodesic
line of the Riemannian space-time $V_{n}$ of the dimensionality $n$ with the
metric tensor $g_{\alpha\beta}\;(\psi)$. That is why if the given gauge
transformation $\psi_{\alpha\beta}\rightarrow\overline{\psi}_{\alpha\beta}$
leaves the equations of motion invariant, then the corresponding
transformation $g_{\alpha\beta}\rightarrow\overline{g}_{\alpha\beta}$ is a
mapping $V\rightarrow\overline{V}$ of the Riemannian spaces leaving geodesic
lines invariant, i.e. it is a geodesic, (projective) mapping. Let us assume
that not only eq. (\ref{Vr_LagrangianThirr}) but also the field equations
contain $\psi_{\alpha\beta}$ only in the form $g_{\alpha\beta}\;(\psi)$, then
it becomes clear that the gauge-invariance of the equations of motion will be
ensured if the field equations are invariant with respect to geodesic mappings
of the Riemannian space $V_{n}$. Thus, if we start from eq.
(\ref{Vr_LagrangianThirr}), then the gravitational field equations as well as
the physical field characteristics must be invariant with respect to geodesic
(projective) mappings of the Riemannian space-time $V_{n}$ with the metric
tensor $g_{\alpha\beta}\;(\psi)$.

The simplest equations of gravitation that can be considered as a realization
of the above idea were proposed in paper \cite{Verozub91}. (From another
wiewpoint). They are given by%

\begin{equation}
B_{\beta\gamma;\alpha}^{\alpha}-B_{\beta\mu}^{\nu}\;B_{\gamma\nu}^{\mu}=0\;,
\label{Myeq}%
\end{equation}
In these equations%

\begin{equation}
B_{\alpha\beta}^{\gamma}=\Pi_{\alpha\beta}^{\gamma}-\overset{\circ}{\Pi
}_{\alpha\beta^{\gamma}}\;, \label{Vr_tens B}%
\end{equation}
where $\Pi_{\alpha\beta}^{\gamma}$ and $\overset{\circ}{\Pi}_{\alpha\beta
}^{\gamma}$ are the Thomas symbols for $V_{n}$ and $E_{n}$,%

\begin{equation}
\Pi_{\alpha\beta}^{\gamma}=\Gamma_{\alpha\beta}^{\gamma}-(n+1)^{-1}\left[
\delta_{\alpha}^{\gamma}\Gamma_{\beta\mu}^{\mu}+\delta_{\beta}^{\gamma}%
\Gamma_{\alpha\mu}^{\mu}\right]  \;,
\end{equation}

\bigskip%
\begin{equation}
\overset{\circ}{\Pi}_{\alpha\beta}^{\gamma}=\overset{\circ}{\Gamma}%
_{\alpha\beta}^{\gamma}-(n+1)^{-1}\left[  \delta_{\alpha}^{\gamma}%
\overset{\circ}{\Gamma}_{\epsilon\beta}^{\epsilon}+\delta_{\beta}^{\gamma
}\overset{\circ}{\Gamma}_{\epsilon\alpha}^{\epsilon}\right]  ,
\end{equation}
$\Gamma_{\alpha\beta}^{\gamma}$ and $\overset{\circ}{\Gamma}_{\alpha\beta
}^{\gamma}$ are the Christoffel symbols in $V_{n}$ and $E_{n},$ respectively.

They are bimetric geodesic invariant equations. Each solution $g_{\alpha\beta
}(x)$ of (\ref{Myeq}) reffers to some coordinate system and is dtermined up to
arbitrary geodesic mappings, which play the role gauge transformation in the
theory under consideration. The physical meaning may have only geodesic
invariant magnitudes, for example, the tensor $B_{\alpha\beta}^{\gamma}.$ At
the covariant gauge conditions $Q_{\alpha}=$ $\Gamma_{\beta\alpha}^{\beta
}-\overset{\circ}{\Gamma}_{\beta\alpha}^{\beta}=0$ eqs. \ref{Myeq} are
equivalent to the system%

\begin{equation}
R_{\alpha\beta}=0 \label{EinsteinEqs}%
\end{equation}
and
\begin{equation}
Q_{\alpha}=0, \label{AditionalCondition}%
\end{equation}
where $R_{\alpha\beta}$ is the Ricci tensor.

\qquad The equations do not contain the functions $\psi_{\alpha\beta}(x)$
explicitly. The simplest way of obtaining equations for such a kind of the
functions $\psi_{\alpha\beta}$ is to set
\begin{equation}
B_{\beta\gamma}^{\alpha}=\nabla^{\alpha}\psi_{\beta\gamma}-(n+1)^{-1}\left(
\delta_{\beta}^{\alpha}\nabla^{\sigma}\psi_{\sigma\gamma}+\delta_{\gamma
}^{\alpha}\nabla^{\sigma}\psi_{\beta\sigma}\right)  , \label{BbyPsi}%
\end{equation}
where $\nabla^{\alpha}$ is the covariant derivative in flat space-time. Then,
at the gauge condition $\nabla^{\sigma}\psi_{\sigma\gamma}=0$ eq. (\ref{Myeq})
are given by
\begin{equation}
\square\psi_{\alpha\beta}-\nabla^{\sigma}\psi_{\alpha\gamma}\,\nabla^{\gamma
}\psi_{\sigma\beta}=0;\ \ \nabla^{\sigma}\psi_{\sigma\gamma}=0,\nonumber
\end{equation}
where $\square$ is the covariant Dalamber operator in pseudo-Euclidean
space-time. It is natural to suppose that with the presence of matter these
equations are given by%

\begin{equation}
\square\psi_{\alpha\beta}=\varkappa(T_{\alpha\beta}+t_{\alpha\beta
});\ \ \;\nabla^{\sigma}\psi_{\sigma\gamma}=0,\; \label{EqPciMatter}%
\end{equation}

where $\varkappa=8\pi G/c^{4},$ $t_{\alpha\beta}=\varkappa^{-1}\nabla^{\sigma
}\psi_{\alpha\gamma}\,\nabla^{\gamma}\psi_{\sigma\beta}$ and $T_{\alpha\beta}$
is the matter tensor of the energy-momentum. Obviously, the equality
\begin{equation}
\nabla^{\beta}(T_{\alpha\beta}+t_{\alpha\beta})=0 \label{ConservLawPsi}%
\end{equation}
is valid. Therefore, the magnitude $t_{\alpha\beta}$ can be interpreted as the
energy-momentum tensor of a gravitational field. \footnote{It should be noted
that, when we introduce it in some way, we cannot be sure aprori that the
equation for $\psi_{\alpha\beta}$ yields all solutions of the equations for
$B_{\beta\gamma}^{\alpha}$. We may introduce a potential $\psi_{\alpha\beta}$
also in another way.} . At the conditions $\nabla^{\sigma}\psi_{\sigma\gamma}=0$%

\begin{equation}
t_{\alpha\beta}=\chi^{-1}B_{\alpha\sigma}^{\gamma}B_{\beta\gamma}^{\sigma
}\label{TensorEnergyGravField}%
\end{equation}

\section{\bigskip Gravitational Energy of a Point Mass}

If the Lagrangian of test particles is invariant under the mapping
$t\rightarrow-t$, the fundamental metric form of space-time $V_{4}$ in the
spherically-symmetric case can be written as
\begin{equation}
ds^{2}=-Adr^{2}-B[d\theta^{2}+\sin^{2}\theta\ d\varphi^{2}]+Cdx^{0}{}%
^{2},\label{Vr_ds}%
\end{equation}
where $A$, $B$ and $C$ are the functions of the radial coordinate $r$. The
general solution of the system (\ref{EinsteinEqs}) --
(\ref{AditionalCondition}) at the conditions
\begin{equation}
\lim\limits_{r\rightarrow\infty}A=1,\;\lim\limits_{r\rightarrow\infty}%
(B/r^{2})=1,\;\lim\limits_{r\rightarrow\infty}C=1.\label{limitConditions}%
\end{equation}
is of the form \cite{Verozub96}%

\begin{equation}
A=(f^{\prime})^{2}(1-\mathcal{Q}/f)^{-1},\;B=f^{2},\;C=1-\mathcal{Q}/f
\label{ABCequal}%
\end{equation}
where
\[
f=(r^{3}+\mathcal{K}^{3})^{1/3}%
\]
$f^{\prime}=df/dr$, $\mathcal{Q}$ and $\mathcal{K}$ are constants.

The equations of the motion of a test particle resulting from Lagrangian
(\ref{Vr_LagrangianThirr}) is given by%

\begin{equation}
\overset{..}{x}^{\alpha}+(\Gamma_{\beta\gamma}^{\alpha}-c^{-1}\Gamma
_{\beta\gamma}^{0}\overset{.}{x}^{\alpha})\overset{.}{x}^{\beta}\overset{.}%
{x}^{\gamma}=0.
\end{equation}
In the nonrelativistic limit $\overset{..}{x}^{r}=-c^{2}\,\Gamma_{00}^{r},$
where $\Gamma_{00}^{r}=C^{\prime}/2A=r^{4}C^{\prime}/f^{4}C.$ Therefore, to
obtain the Newton gravity law it should be supposed that at large $r$ the
function $f\approx r$ and $\mathcal{Q}=r_{g}=2G\,M/c^{2}$ is the classical
Schwarzshild radius.

We can also argue that the constant $\mathcal{K}=r_{g}.$ Indeed, consider the
00-component of the first of eq. (\ref{EqPciMatter}). Let us set
$T_{\alpha\beta}=\rho c^{2}u_{\alpha}u_{\beta},$ where $\rho$ is the matter
density and $u_{\alpha}$ is the 4-velocity of matter points. At the small
macroscopic velocities of the matter we can set $u_{0}=1$ and $u_{i}=0$.
Therefore, the equation is of the form
\begin{equation}
\square\psi_{\alpha\beta}=\chi(\rho c^{2}+t_{00})
\end{equation}
where $\chi=4\pi G/c^{4}$ and $t_{00}$ is the 00-component of the tensor
(\ref{TensorEnergyGravField}). Let us find the energy of a gravitational field
of the point mass $M$ as the following integral in the pseudo-Euclidean
space-time
\begin{equation}
\mathcal{E}=\int t_{00}dV,
\end{equation}
resulting from the above solution, where $dV$ is the volume element. In the
Newtonian theory this integral is divergent. In our case we have:
\begin{equation}
\mathcal{E}=\frac{\mathcal{Q}}{\mathcal{K}}M\,c^{2}\label{Vr_energyfinally}%
\end{equation}
We arrive at the conclusion that at $\mathcal{K}\neq0$ the energy of the point
mass is finite and \ at $\mathcal{K}=\mathcal{Q}$ the rest energy of the point
particle in full is caused by its gravitational field: $\mathcal{E}=M\,c^{2}$

The spacial components of the vector $P_{\alpha}=t_{0\alpha}$ are equal to
zero. Due to these facts we assume in the present paper that $\mathcal{K}%
=\mathcal{Q}=r_{g}$ and consider the solution (\ref{ABCequal}) in the
spherical coordinate system at the used gauge condition as a basis for the
subsequent analysis.

\section{Acceleration of the Universe as a Consequence of Gravitation Properties}

The analysis of the recent observations data gives evidence that the
deceleration parameter $q_{o}=-\overset{..}{a}(t)$ $a(t)/\overset{.}{a}(t)$ (
$a$ is the scale factor) is negative at the moment (\cite{Riess} ,
\cite{Perlmuter}) . It means that $\overset{..}{a}>0$ i.e. the expansion is
accompanied with acceleration, while according to classical insights the
gravity force must retard the expansion.

Equations (\ref{Myeq}) were succesfuly testified by the classical tests and
binary pulsar PSR1913+16 \cite{Verozub96}, \cite{VerKoch2000}. The motion of
test particles in the spherically - symmetric gravitational field differs very
little from that in general relativity at distances from the center $r\gg
r_{g}.$ However, they are completely different at $r\leq r_{g}.$ The solution
of (\ref{Myeq}) has no the event horizon and physical singularity in the
center. The gravitational force (as the mass multiplied by the acceleration)
affecting escape particles is repulsive from $r=0$ up to distances of the
order of $r_{g}$. The observed radius $R_{U}$ of the Universe is about
$10^{27}cm.$ and the observed mass $M_{U}$ is $10^{56}\div10^{57}g$ , so the
magnitude $2GM_{U}/c^{2}$ is of the order of $R_{U}$ . For this reason, we can
expect some manifastation of the repulsive force for distant objects.

Consider in flat space-time a simle model of an expanding selfgraviting
homogeneous dust - ball with the sizes of the observed Universe. Acording to
\cite{Verozub91}, the \qquad equation of the motion of a test particle in the
spherically-symmetric field are given by%

\begin{equation}
{\overset{\cdot}{r}}^{2}=(c^{2}C/A)[1-(C/\overline{E})(1+r_{g}^{2}\overline
{J}^{2}/B)], \label{EqsMotionTestPart1}%
\end{equation}%
\begin{equation}
\overset{\cdot}{\varphi}=c\;C\overline{J}r_{g}/(B\overline{E})
\label{EqsMotionTestPart2}%
\end{equation}
where $(r,\varphi,\theta)$ are the spherical coordinates ($\theta$ is supposed
to be equal to $\pi/2$), $\overset{\cdot}{r}=dr/dt$, $\overset{\cdot}{\varphi
}=d\varphi/dt$ , $\overline{E}=E/(mc^{2})$, $\overline{J}=J/(amc)$, $E$ is the
particle energy, $J$ is the angular momentum.

Consequently, the radial velocity of the specks of dust of the ball surface as
a function of its radius $R$ is given by
\begin{equation}
V^{2}=\frac{c^{2}C}{A}\left[  1-\frac{C}{\overline{E}^{2}}\right]  ,
\label{eq1_motion}%
\end{equation}

where , $V=\dot{R}=dR/dt$,%

\begin{equation}
C=1-\frac{1}{\overset{\_}{f}};\ \ A=\frac{\overset{\_}{r}^{4}}{\overset{\_}%
{f}^{4}}C^{-1};\ \ ,\overset{\_}{f}=(1+\overset{\_}{r}^{3})^{1/3}%
\end{equation}
and the dependence of the acceleration $\overset{\cdot}{V}=dV/dt$ on the
radius $R$ is given by%

\begin{equation}
\overset{\cdot}{V}=VV^{\prime}%
\end{equation}

\begin{figure}
[ptb]
\begin{center}
\includegraphics[
height=5.0918cm,
width=7.0951cm
]%
{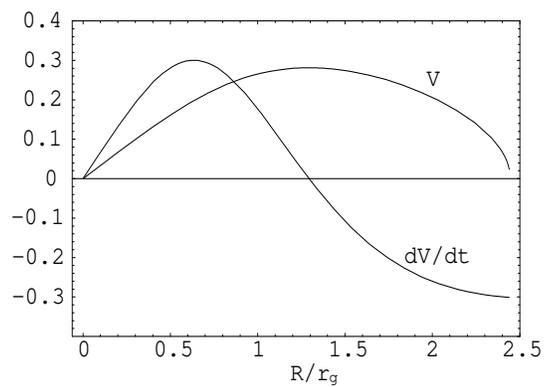}%
\caption{The radial velocity on the ball surface vs.  the radius $R$ at
$\overline{E}^{2}<1.$}%
\label{fig3.eps}%
\end{center}
\end{figure}

\begin{figure}
[ptb]
\begin{center}
\includegraphics[
height=5.0918cm,
width=7.7958cm
]%
{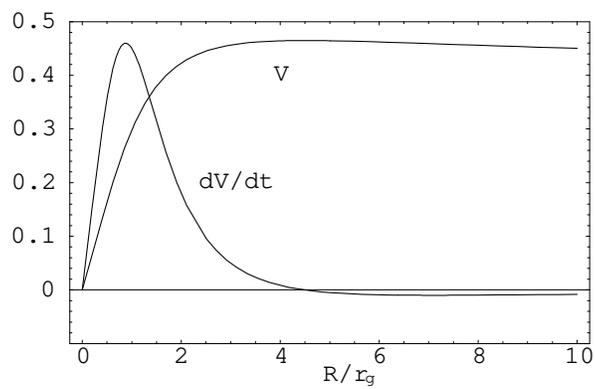}%
\caption{The radial velocity and acceleration on the ball surface vs. the 
radius $R$ at $\overline{E}^{2}<1.$}%
\label{fi4.eps}%
\end{center}
\end{figure}

Figs.\ref{fig3.eps} and \ref{fi4.eps} show the plot of the velocity and the
acceleration (arbitrary units) as the function of the radius at $\overline
{E}^{2}=0.60<1$ and $\overline{E}^{2}=1.20>1$ .%

\bigskip%

It follows from the figures that the acceleration  is negative at f
$R/r_{g}\gg1$, and is positive if $R/r_{g}$ is of the order of $r_{g}$ or less
than that. For example, if the radius $R=R_{U}$ and the matter density
$\rho=2\cdot10^{=28}g/cm^{3}$ , the value of $R_{U}/r_{g}=0.9<1$ and the
acceleration is equal to $1\cdot10^{-8}cm/s^{2}$ which is half as large as
this magnitude resulting from the value of $q_{0}=-1$ that was found in
\cite{Riess}.

In paper \cite{VerKoch01} the Riess et al. results were studied in detail in
view of the model above. A good compliance was found.

\section{Conclusion}

The key reason preventing a correct inclusion of the Einstein theory of
gravitation in the interactions unification is that gravity is identified with
space-time curvature. It is also a cause of such unsolved problems of the
theory as an operational definition of the observables, the energy - momentum
tensor problem and gravity quantization. So, if the analysed above possibility
really takes place in nature, then it will remove an isolation of the
geometrical gravitational theories from the theories of other fields.

\end{document}